*Review*

# The Status of Cosmic Topology after Planck Data

Jean-Pierre Luminet

LAM (Laboratoire d'Astrophysique de Marseille), CNRS UMR 7326, F-13388 Marseille, France ; LUTH (Observatoire de Paris), CNRS UMR 8102, F-92195 Meudon, France ; CPT (Centre de Physique Théorique), CNRS UMR 7332, F-13288 Marseille, France; jean-pierre.luminet@lam.fr

**Abstract:** In the last decade, the study of the overall shape of the universe, called Cosmic Topology, has become testable by astronomical observations, especially the data from the Cosmic Microwave Background (hereafter CMB) obtained by WMAP and Planck telescopes. Cosmic Topology involves both global topological features and more local geometrical properties such as curvature. It deals with questions such as whether space is finite or infinite, simply-connected or multi-connected, and smaller or greater than its observable counterpart. A striking feature of some relativistic, multi-connected small universe models is to create multiples images of faraway cosmic sources. While the last CMB (Planck) data fit well the simplest model of a zero-curvature, infinite space model, they remain consistent with more complex shapes such as the spherical Poincaré Dodecahedral Space, the flat hypertorus or the hyperbolic Picard horn. We review the theoretical and observational status of the field.

## 1. Introduction

The idea that the universe might have a non-trivial topology and, if sufficiently small in extent, display multiple images of faraway sources, was first discussed in 1900 by Karl Schwarzschild (see [1] for reference and English translation). With the advent of Einstein's general relativity theory (see, e.g., the recent historical overview in [2]) and the discoveries of non-static universe models by Friedmann and Lemaître in the decade 1922–1931, the face of cosmology definitively changed. While Einstein's cosmological model of 1917 described space as the simply-connected, positively curved hypersphere $S^3$, de Sitter in 1917 and Lemaître in 1927 used the multi-connected projective sphere $P^3$ (obtained by identifying opposite points of $S^3$) for describing the spatial part of their universe models. In 1924, Friedmann [3] pointed out that Einstein's equations are not sufficient for deciding if space is finite or infinite: Euclidean and hyperbolic spaces, which in their trivial (*i.e.*, simply-connected) topology are infinite in extent, can become finite (although without an edge) if one identifies different points—an operation which renders the space multi-connected. This opens the way to the existence of phantom sources, in the sense that at a single point of space an object coexists with its multiple images. The whole problem of cosmic topology was thus posed, but as the cosmologists of the first half of 20th century had no experimental means at their disposal to measure the topology of the universe, the vast majority of them lost all interest in the question. A revival of interest in multi-connected cosmologies arose in the 1970s, under the lead of theorists who investigated several kinds of topologies (see [4] for an exhaustive review and references, in which the term "Cosmic Topology" was coined). However, most cosmologists either remained completely ignorant of the possibility, or regarded it as unfounded, until the 1990s when data on the CMB provided by space telescopes gave access to the largest possible volume of the observable universe. Since then, hundreds of articles have considerably enriched the field of theoretical and observational cosmology.

## 2. A Theoretical Reminder

At very large scale our Universe seems to be correctly described by one of the Friedmann-Lemaître (hereafter FL) models, namely homogeneous and isotropic solutions of Einstein's equations, of which the spatial sections have constant curvature. They fall into three general classes, according to the sign of curvature. In most cosmological studies, the spatial topology is assumed to be that of the corresponding simply-connected space: the hypersphere, Euclidean space or 3D-hyperboloid, the first being finite and the other two infinite. However, there is no particular reason for space to have a trivial topology: the Einstein field equations are local partial differential equations which relate the metric and its derivatives at a point to the matter-energy contents of space at that point. Therefore, to a metric element solution of Einstein field equations, there are several, if not an infinite number, of compatible topologies, which are also possible models for the physical universe. For example, the hypertorus $T^3$ and the usual Euclidean space $E^3$ are locally identical, and relativistic cosmological models describe them with the same FL equations, even though the former is finite in extent while the latter is infinite. Only the boundary conditions on the spatial coordinates are changed. The multi-connected FL cosmological models share exactly the same kinematics and dynamics as the corresponding simply-connected ones; in particular, the time evolutions of the scale factor are identical.

In FL models, the curvature of physical space (averaged on a sufficiently large scale) depends on the way the total energy density of the universe may counterbalance the kinetic energy of the expanding space. The normalized density parameter $\Omega_0$, defined as the ratio of the actual energy density to the critical value that strictly Euclidean space would require, characterizes the present-day contents (matter, radiation and all forms of energy) of the universe. If $\Omega_0$ is greater than 1, then the space curvature is positive and the geometry is spherical; if $\Omega_0$ is smaller than 1, the curvature is negative and geometry is hyperbolic; eventually $\Omega_0$ is strictly equal to 1 and space is locally Euclidean (currently said flat, although the term can be misleading).

Independently of curvature, a much discussed question in the history of cosmology (and also philosophy) is to know whether space is finite or infinite in extent. Of course no physical measure can ever prove that space is infinite, but a sufficiently small, finite space model could be testable. Although the search for space topology does not necessarily solve the question of finiteness, it provides many multi-connected universe models of finite volume.

The effect of a non-trivial topology on a cosmological model is equivalent to considering the observed space as a simply-connected 3D-slice of space-time, known as the "universal covering space" (hereafter UC) being filled with repetitions of a given shape, the "fundamental domain", which is finite in some or all directions, for instance a convex polyhedron; by analogy with the two-dimensional case, we say that the fundamental domain tiles the UC space.

There is a subgroup of isometries acting on the UC which produces its tiling by these copies (for such group action, see the basic works [5,6]). Physical fields repeat their configuration in every copy and thus can be viewed as defined on the UC space, but subject to periodic boundary conditions, which are the matching rules between neighbouring tiles. The copies around a fixed one carry the multiple images of objects from the cosmos. By analogy with crystallography, the UC plays the role of the macroscopic crystal, the cosmos plays the role of the fundamental unit (see Figure 1). But in contrast to crystallography, the UC in topology can be Euclidean, spherical or hyperbolic. For the flat and hyperbolic geometries, there are infinitely many copies of the fundamental domain; for the spherical geometry with a finite volume, there are a finite number of tiles.

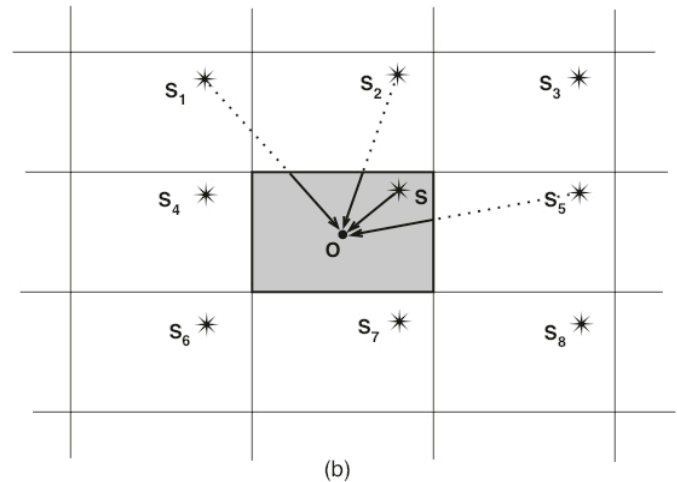

(b)

**Figure 1.** The Illusion of the Universal Covering Space. In the case of a 2D torus space, the fundamental domain, which represents real space, is the interior of a rectangle, whose opposite edges are identified. The observer O sees rays of light from the source S coming from several directions. He has the illusion of seeing distinct sources $S_1$, $S_2$, $S_3$, *etc.*, distributed along a regular canvas which covers the UC space—an infinite plane.

There are seventeen multi-connected Euclidean spaces (for an exhaustive study, see [7]), the simplest of which being the hypertorus $T^3$, whose fundamental domain is a parallelepiped of which opposite faces are identified by translations. Seven of these spaces have an infinite volume, ten are of finite volume, six of them being orientable hypertori. All of them could correctly describe the spatial part of the flat universe models, as they are consistent with recent observational data which constrain the space curvature to be very close to zero. Note that in current inflationary scenarios for the big bang, one can always have a nearly flat universe at present without fine-tuning the initial value of the spatial curvature, while considering exactly flat models corresponds to fine-tuning the initial curvature to be strictly zero.

In spaces with non-zero curvature, the presence of a length scale (the curvature radius) precludes topological compactification at an arbitrary scale. The size of the space must now reflect its curvature, linking topological properties to the total energy density $\Omega_0$. All spaces of constant positive curvature are finite whatever be their topology. The reason is that the universal covering space—the simply-connected hypersphere $S^3$—is itself finite. There is a countable infinity of spherical spaceforms (for a complete classification, see [8]), but there is only a finite set of "well-proportioned" topologies, *i.e.*, those with roughly comparable sizes in all directions, which are of a particular interest for cosmology. As a now celebrated example, let us mention the Poincaré Dodecahedral Space (hereafter PDS), obtained by identifying the opposite pentagonal faces of a regular spherical dodecahedron after rotating by 36° in the clockwise or counterclockwise direction around the axis orthogonal to the face, depending on which handedness the physical nature favors [9], see Figure 2. Its volume is 120 times smaller than that of the hypersphere with the same curvature radius.

Eventually there is also an infinite (but non countable) number of hyperbolic manifolds, with finite or infinite volumes. Their classification is not well understood, but the volumes of the compact hyperbolic space forms are bounded below by $V = 0.94271$ (in units of the curvature radius), which correspond to the so-called Weeks manifold.

The computer program *CurvedSpaces* [10] is especially useful to depict the rich structure of multi-connected manifolds with any curvature.

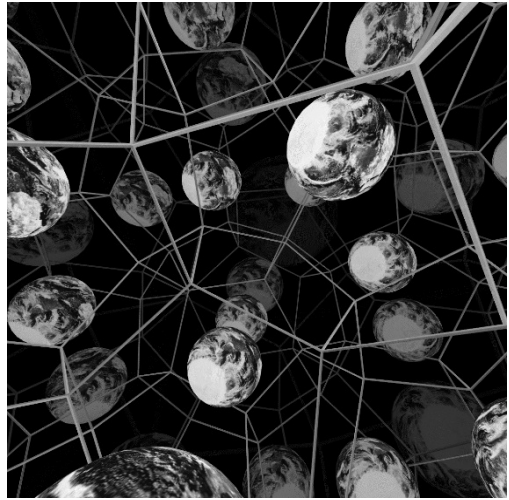

**Figure 2.** The Poincaré Dodecahedral Space can be described as the interior of a spherical ball whose surface is tiled by 12 curved regular pentagons. When one leaves through a pentagonal face, one returns to the ball through the opposite face after having turned by 36°. As a consequence, the space is finite but without boundaries, therefore one can travel through it indefinitely. One has the impression of living in a UC space 120 times larger, paved with dodecahedra that multiply the images like a hall of mirrors. The return of light rays that cross the walls produces optical mirages: a single object has several images. This numerical simulation calculates the closest phantom images of the Earth, which would be seen in the UC space (Image courtesy of J. Weeks).

## 3. Probing Cosmic Topology

The observable universe is the interior of a sphere centered on the observer and whose radius is that of the cosmological horizon—roughly the radius of the last scattering surface (hereafter LSS), presently estimated at 14.4 Gpc. Cosmic Topology aims to describe the shape of the *whole* universe. One could think that the whole universe is necessarily greater than the observable one, as it would obviously be the case if space was infinite, for instance the simply-connected flat or hyperbolic space. Then the observable universe would be an infinitesimal patch of the whole universe and, although it has long been the standard "mantra" of many theoretical cosmologists, this is not and will never be a testable hypothesis.

The whole universe can also be finite (without an edge), e.g., a hypersphere or a closed multi-connected space, but greater than the observable universe. In that case, one easily figures out that if whole space widely encompasses the observable one, no signature of its finiteness will show in the experimental data. But if space is not too large, or if space is not globally homogeneous (as is permitted in many space models with multi-connected topology), and if the observer occupies a special position, some imprints of the space finiteness could be observable.

Surprisingly enough, the whole space could be smaller than the observable universe, due to the fact that space can be both multi-connected, have a small volume and produce topological lensing. This is the only case where there are a lot of testable possibilities, whatever the curvature of space.

The present observational constraints on the $\Omega_0$ parameter favor a spatial geometry that is nearly flat with a 0.4% margin of error [11]. Note that the constraints on the curvature parameter can be looser if we consider a general form of dark energy (not the cosmological constant), which leaves rooms to consider positively or negatively curved cosmological models that are usually regarded as being excluded. However, even with the curvature so severely constrained by cosmological data, there are still possible multi-connected topologies that support positively curved, negatively curved, or flat metrics. Sufficiently "small" universe models would generate multiple images of some light sources, in such a way that the hypothesis of multi-connected topology can be tested by astronomical observations. The smaller the space, the easier it is to observe the multiple images of luminous sources in the sky (generally not seen at the same age, except for the CMB spots). Note, however, the coincidence problem that occurs in order to get an observable non-trivial topology: for flat space, we need to have the topology scale length near the horizon scale, while for curved spaces, the curvature radius needs to be near the horizon scale. However, there are so many other, non-explained coincidence problems in standard cosmology that it should not deviate our attention from the possibility of a detectable topology.

How do the present observational data constrain the possible multi-connectedness of the universe and, more generally, what kinds of tests are conceivable (see [12] for a non-technical book about all aspects of topology and its applications to cosmology)?

Different approaches have been proposed for extracting information about the topology of the universe from experimental data. One approach is to use the 3D distribution of astronomical objects such as galaxies, quasars and galaxy clusters: if the whole universe is finite and small enough, we should be able to see "all around" it because the photons might have crossed it once or more times. In such a case, any observer might recognize multiple images of the same light source, although distributed in different directions of the sky and at various redshifts, or to detect specific statistical properties in the distribution of faraway sources. Various methods of "cosmic crystallography", initially proposed in [13], have been widely developed by other groups ([14] and references therein). However, for plausible small universe models, the first signs of topological lensing would appear only at pretty high redshift, say $z \approx 2$. The main limitation of cosmic crystallography is that the presently available catalogs of observed sources at high redshift are not complete enough to perform convincing tests for topology. For instance, looking for nontoroidal topological lensing, [15] applied the crystallographic method to the SDSS quasar sample; though they found no robust signature, cosmological interpretation of the result was prohibited by the data incompleteness and by the uncertainty in quasar physics. On the other hand, [16] proposed to use deep surveys of distant (at redshifts z ~ 6) starburst galaxies for an independent test of the cubic hypertorus model. Their calculation showed that even photometric redshifts would suffice in this purpose, which makes their strategy a realistic and interesting one.

The other approach uses the 2D CMB maps (for a review, [17]). The last scattering surface (LSS) from which the CMB is released represents the most distant source of photons in the universe, hence the largest scales with which the topology of the universe can be probed.

The idea that a small universe model could lead to a suppression of power on large angular scales in the fluctuation spectrum of the CMB had been proposed in the 1980s [18]: in some way, space would be not big enough to sustain long wavelengths. After the release of COBE data in 1992 and the higher resolution and sensitivity of WMAP (2003), there were indeed indications of low power on large scales which could have had a topological origin, and many authors used it to constrain the models. The best fits between theoretical power spectra computed for various topologies and the observed one were obtained with the positively curved Poincaré Dodecahedral Space [9,19,20] and the flat hypertorus [21]. In addition, it was shown [22] that the low-order multipoles tended to be relatively weak in "well-proportioned" spaces, *i.e.*, whose dimensions are approximately equal in all directions. Some globally inhomogeneous topologies can also reproduce the large-angle CMB power suppression if the location of the observer is so adjusted that his fundamental domain becomes well-proportioned [23]. However, this possibility was not borne out by detailed real- and harmonic-space analyses in two dimensions, so that the arguments based on the angular power spectrum and favoring small universe models failed [24]. In any case, to gain all the possible information from the correlations of CMB anisotropies, one has to consider the full covariance matrix rather than just the power spectrum.

Indeed the main imprint of a non trivial topology on the CMB is well-known in the case when the characteristic topological length scale (called the injectivity radius) is smaller than the radius of the LSS: the crossings of the LSS with its topological images give rise to pairs of matched circles of equal radii, centered at different points on the CMB sky, and exhibiting correlated patterns of temperature variations [25], see Figure 3. For instance, the PDS model predicts six pairs of antipodal circles with an angular radius comprised between 5° and 55° (sensitively depending on the cosmological parameters) and a relative phase of 36°.

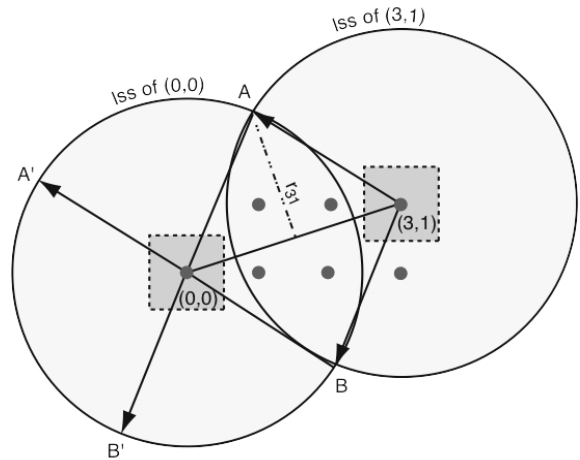

**Figure 3.** The circles-in-the-sky method is illustrated here in a 2D torus space. The fundamental domain is a square (with a dotted outline), all of the red points are copies of the same observer. The two large circles (which are normally spheres in a three-dimensional space) represent the last scattering surfaces (LSS) centered on two copies of the same observer. One is in position (0,0), its copy is in position (3,1) in the universal covering space. The intersection of the circles is made up of the two points A and B (in three dimensions, this intersection is a circle). The observers (0,0) and (3,1), who see the two points (A,B) from two opposite directions, are equivalent to a unique observer at (0,0) who sees two identical pairs (A,B) and (A′,B′) in different directions. In three dimensions, the pairs of points (A,B) and (A′,B′) become a pair of identical circles, whose radius $r_{31}$ depends on the size of the fundamental domain and the topology.

## 4. Results and Discussion

Such "circles-in-the-sky" searches have been looked for in WMAP maps by several groups, using various statistical indicators and massive computer calculations, and interpreting their results differently. Some authors [26] claimed that most of non-trivial topologies, including PDS and $T^3$, were ruled out: they searched for antipodal or nearly antipodal pairs of circles in the WMAP data and found no such circles. However, their analysis could not be applied to more complex topologies, for which the matched circles deviate strongly from being antipodal. On the other hand, other groups claimed to have found hints of multi-connected topology, using different statistical indicators [27–29].

Most studies have emphasized searches for fundamental domains with antipodal correlations. The search for matched circle pairs that are not back-to-back has nevertheless been carried out recently, with no obvious topological signal appearing in the seven-year WMAP data [30]. The statistical significance of such results still has to be clarified. In any case, a lack of nearly matched circles does not exclude a multi-connected topology on scale less than the horizon radius: detectable topologies may produce circles of small radii which are statistically hard to detect and current analysis of CMB sky maps could have missed even antipodal matching circles, because various effects may damage or even destroy the temperature matching.

Other methods for experimental detection of non-trivial topologies have thus been proposed and used to analyze the experimental data, such as the multipole vectors and the likelihood (Bayesian) method. The latter ameliorates some of the spoiling effects of the temperature correlations such as the integrated Sachs-Wolfe and Doppler contributions [31].

The most up-to-date study based on CMB temperature correlations used the Planck 2013 intensity data [32]. In that work, they applied two techniques: first, a direct likelihood calculation of (a very few) specific topological models; second, a search for the expected repeated "circles in the sky", calibrated by simply-connected simulations. Both of these showed that the scale of any possible topology must exceed roughly the comoving distance to the LSS, $\chi_{rec}$. For the cubic torus, they found that the radius of the largest sphere inscribed in the topological fundamental domain must be $R_i > 0.92\ \chi_{rec}$. The matched-circle limit on topologies predicting back-to-back circles larger than 15° in angular radius and assuming that the relative orientation of the fundamental domain and mask allows its detection was $R_i > 0.94\ \chi_{rec}$ at the 99% confidence level.

Finally, it is now widely understood that the polarization of the CMB can provide a lot of additional informations for reconstructing the cosmological model. Riazuelo *et al.* [33] were the first to show how the polarization could also be used to put additional constraints on space topology and a little bit tighter than those coming from temperature intensity. Maps of CMB polarization from the 2015 release of Planck data [34] provided the highest-quality full-sky view of the LSS available to date. However their study specialized only to flat spaces with cubic toroidal ($T^3$) and slab ($T^1$) topologies. These searches yield no detection of a compact topology with a scale below the diameter of the LSS. More precisely, $R_i > 0.97\ \chi_{rec}$ for the $T^3$ cubic torus and $R_i > 0.56\ \chi_{rec}$ for the $T^1$ slab.

## 5. Conclusions

The overall topology of the universe has become an important concern in astronomy and cosmology. Even if particularly simple and elegant models such as the PDS and the hypertorus are now claimed to be ruled out at a subhorizon scale, many more complex models of multi-connected space cannot be eliminated as such. In addition, even if the size of a multi-connected space is larger (but not too much) than that of the observable universe, we could still discover an imprint in the CMB, even while no pair of circles, much less ghost galaxy images, would remain. The topology of the universe could therefore provide information on what happens outside of the cosmological horizon [35].

Whatever the observational constraints, a lot of unsolved theoretical questions remain. The most fundamental one is the expected link between the present-day topology of space and its quantum origin, since classical general relativity does not allow for topological changes during the course of cosmic evolution. Theories of quantum gravity should allow us to address the problem of a quantum origin of space topology. For instance, in quantum cosmology, the question of the topology of the universe is completely natural. Quantum cosmologists seek to understand the quantum mechanism whereby our universe (as well as other ones in the framework of multiverse theories) came into being, endowed with a given geometrical and topological structure. We do not yet have a correct quantum theory of gravity, but there is no sign that such a theory would *a priori* demand that the universe have a trivial topology. Wheeler first suggested that the topology of space-time might fluctuate at a quantum level, leading to the notion of a space-time foam [36]. Additionally, some simplified solutions of the Wheeler-de Witt equations for quantum cosmology show that the sum over all topologies involved in the calculation of the wavefunction of the universe is dominated by spaces with small volumes and multi-connected topologies [37]. In the approach of brane worlds in string/M-theories, the extra-dimensions are often assumed to form a compact Calabi-Yau manifold; in such a case, it would be strange that only the ordinary, large dimensions of our 3-brane would not be compact like the extra ones. However, still at an early stage of development, string quantum cosmology can only provide heuristic indications on the way multi-connected spaces would be favored.

**Acknowledgments:** The author thanks the anonymous referees for suggesting additional references and improvements of the original manuscript.

**Conflicts of Interest:** The author declares no conflict of interest.

## References

1. Starkman, G.D. Topology and Cosmology. *Class. Quantum Gravity* **2008**, *15*, 2529–2538.
2. Iorio, L. Editorial for the Special Issue 100 Years of Chronogeometrodynamics: The Status of the Einstein's Theory of Gravitation in Its Centennial Year. *Universe* **2015**, *1*, 38–81.

3. Friedmann, A. Über die Möglichkeit einer Welt mit konstanter negativer Krümmung des Raumes. *Zeitsfricht Phys.* **1924**, *21*, 326–332. (English Translation in Bernstein, J.; Feinberg, G. *Cosmological Constants. Papers in Modern Cosmology*; Columbia University Press: New York, NY, USA, 1986; pp. 59–65.)
4. Lachièze-Rey, M.; Luminet, J.-P. Cosmic topology. *Phys. Rep.* **1995**, *254*, 135–214.
5. Wolf, J.A. *Spaces of Constant Curvature*; Publish or Perish Inc.: Wilmington, DC, USA, 1984.
6. Kramer, P. Topology of Platonic Spherical Manifolds: From Homotopy to Harmonic Analysis. *Symmetry* **2015**, *7*, 305–326.
7. Riazuelo, A.; Weeks, J.; Uzan, J.-P.; Lehoucq, R.; Luminet, J.-P. Cosmic microwave background anisotropies in multiconnected flat spaces. *Phys. Rev. D* **2004**, *69*, 103518.
8. Gausmann, E.; Lehoucq, R.; Luminet, J.-P.; Uzan, J.-P.; Weeks, J. Topological lensing in spherical spaces. *Class. Quantum Gravity* **2001**, *18*, 5155–5186.
9. Luminet, J.-P.; Weeks, J.; Riazuelo, A.; Lehoucq, R.; Uzan, J.-P. Dodecahedral space topology as an explanation for weak wide-angle temperature correlations in the cosmic microwave background. *Nature* **2003**, *425*, 593–595.
10. Geometrygames. Available online: http://www.geometrygames.org (accessed on 8 January 2016).
11. Hinshaw, G.; Larson, D.; Komatsu, E.; Spergel, D. N.; Bennett, C. L.; Dunkley, J.; Nolta, M. R.; Halpern, M.; Hill, R. S.; Odegard, N; *et al*. Nine-year WMAP Observations: Cosmological Parameter Results. *Astrophys. J. Suppl.* **2013**, *208*, 19.
12. Luminet, J.-P. *The Wraparound Universe*; AK Peters: New York, NY, USA, 2008.
13. Lehoucq, R.; Lachièze-Rey, M.; Luminet, J.-P. Cosmic crystallography. *Astron. Astrophys.* **1996**, *313*, 339–346.
14. Fujii, H.; Yoshii, Y. An improved cosmic crystallography method to detect holonomies in flat spaces. *Astron. Astrophys.* **2011**, *529*, A121.
15. Fujii, H.; Yoshii, Y. A Search for Nontoroidal Topological Lensing in the Sloan Digital Sky Survey Quasar Catalog. *Astrophys. J.* **2013**, *773*, 152–160.
16. Roukema, B.; France, M.; Kazimierczak, T.; Buchert, T. Deep redshift topological lensing: Strategies for the $T^3$ candidate. *Mon. Not. R. Astron. Soc.* **2014**, *437*, 1096–1108.
17. Levin, J. Topology and the cosmic microwave background. *Phys. Rep.* **2002**, *365*, 251–333.
18. Fagundes, H.V. The Quadrupole Component of the Relic Radiation in a Quasi-Hyperbolic Cosmological Model. *Astrophys. Lett.* **1983**, *23*, 161.
19. Caillerie, S.; Lachièze-Rey, M.; Luminet, J.-P. A new analysis of the Poincaré dodecahedral space model. *Astron. Astrophys.* **2007**, *476*, 691–696.
20. Roukema, B.F.; Bulinski, Z.; Gaudin, N.E. Poincaré dodecahedral space parameter estimates. *Astron. Astrophys.* **2008**, *492*, 657–673.
21. Aurich, R. A spatial correlation analysis for a toroidal universe. *Class. Quantum Gravity* **2008**, *25*, 225017.
22. Weeks, J.; Luminet, J.-P.; Riazuelo, A.; Lehoucq, R. Well-proportioned universes suppress the cosmic microwave background quadrupole. *Mon. Not. R. Astron. Soc.* **2004**, *352*, 258–262.
23. Aurich, R.; Lustig, S. How well proportioned are lens and prism spaces? *Class. Quantum Gravity* **2012**, *29*, 175003.
24. Bielewicz, P.; Riazuelo, A. The study of topology of the Universe using multipole vectors. *Mon. Not. R. Astron. Soc.* **2009**, *396*, 609–623.
25. Cornish, N.J.; Spergel, D.N.; Starkman, G.D. Circles in the sky: finding topology with the microwave background radiation. *Class. Quantum Gravity* **1998**, *15*, 2657–2670.
26. Cornish, N.J.; Spergel, D.N.; Starkman, G.D.; Komatsu, E. Constraining the Topology of the Universe. *Phys. Rev. Lett.* **2004**, *92*, 201302.
27. Roukema, B.; Lew, B.; Cechowska, M.; Marecki, A.; Bajtlik, S. A Hint of Poincaré Dodecahedral Topology in the WMAP First Year Sky Map. *Astron. Astrophys.* **2004**, *423*, 821–831.
28. Lew, B.; Roukema, B. A test of the Poincaré dodecahedral space topology hypothesis with the WMAP CMB data. *Astron. Astrophys.* **2008**, *482*, 747–753.
29. Aurich, R.; Lustig, S.; Steiner, F. The circles-in-the-sky signature for three spherical universes. *Mon. Not. R. Astron. Soc.* **2006**, *369*, 240–248.
30. Vaudrevange, P.; Starkman, G.; Cornish, N.; Spergel, D. Constraints on the topology of the Universe: Extension to general geometries. *Phys. Rev. D* **2012**, *86*, 083526.

31. Kunz, M.; Aghanim, N.; Cayon, L.; Forni, O.; Riazuelo, A.; Uzan, J.-P. Constraining topology in harmonic space. *Phys. Rev. D* **2006**, *73*, 023511.
32. Ade, P. A. R.; Aghanim, N.; Armitage-Caplan, C.; Arnaud, M.; Ashdown, M.; Atrio-Barandela, F.; Aumont, J.; Baccigalupi, C.; Banday, A. J.; Barreiro, R. B.; *et.al*;. Planck 2013 results XXVI. Background geometry and topology of the Universe. *Astron. Astrophys.* **2014**, *571*, A26.
33. Riazuelo, A.; Caillerie, S.; Lachièze-Rey, M.; Lehoucq, R.; Luminet, J.-P. Constraining Cosmic Topology with CMB Polarization. **2006**, arXiv:astro-ph/0601433.
34. Ade, P. A. R.; Aghanim, N.; Arnaud, M.; Ashdown, M.; Aumont, J.; Baccigalupi, C.; Banday, A. J.; Barreiro, R. B.; Bartolo, N.; Battaner, E.; *et.al*. Planck 2015 results XVIII. Background geometry and topology. **2015**. arXiv:1502.01593.
35. Fabre, O.; Prunet, S.; Uzan, J.-P. Topology beyond the horizon: how far can it be probed? *Phys. Rev. D* **2015**, *92*, 04003.
36. Wheeler, J. On the nature of quantum geometrodynamics. *Ann. Phys.* **1957**, *2*, 604–614.
37. e Costa, S.S.; Fagundes, H.V. On the Birth of a Closed Hyperbolic Universe. *Gen. Relativ. Gravit.* **2001**, *33*, 1489–1494.